\documentclass[a4paper,11pt]{article}
\usepackage{graphicx} 

\pdfoutput=1 

\usepackage{jcappub} 

\usepackage[T1]{fontenc} 

\usepackage[utf8]{inputenc}
\usepackage{amsmath,mathtools}

\usepackage{tensor}
\usepackage{amssymb}
\usepackage{graphicx}
\usepackage{physics}
\usepackage{lipsum}  
\usepackage{bigints}
\usepackage{verbatim}

\usepackage{braket}
\usepackage{cancel}
\usepackage{booktabs}
\usepackage{xcolor}
\usepackage{hyperref}
\usepackage{graphicx,subcaption}

\newcommand{\roughly}[1]{\mathrel{\raise.3ex\hbox{$#1$\kern-0.85em
\lower1ex\hbox{$\sim$}}}}
\newcommand{\lsim}{\roughly<}
\newcommand{\gsim}{\roughly>}

\newcommand{\mfa}{{\mathfrak a}}

\newcommand{\bfk}{{\mathbf{k}}}
\newcommand{\bfr}{{\mathbf{r}}}

\newcommand{\cH}{{\cal H}}
\newcommand{\cL}{{\cal L}}
\newcommand{\cO}{{\cal O}}

\newcommand{\ssB}{{\scriptscriptstyle B}}

\newcommand{\ssH}{{\scriptscriptstyle H}}
\newcommand{\ssI}{{\scriptscriptstyle I}}
\newcommand{\ssN}{{\scriptscriptstyle N}}

\newcommand{\ssR}{{\scriptscriptstyle R}}

\newcommand{\exd}{{\rm d}}
\newcommand{\pref}[1]{{(\ref{#1})}}

\newcommand{\nn}{\nonumber}
\newcommand{\be}{\begin{equation}}
\newcommand{\ee}{\end{equation}}
\newcommand{\bea}{\begin{eqnarray}}
\newcommand{\eea}{\end{eqnarray}}

\newcommand{\sss}{\scriptscriptstyle}
\newcommand{\ol}{\overline}
\newcommand{\EW}{{\sss EW}}
\newcommand{\DE}{{\sss DE}}
\newcommand{\KK}{{\sss KK}}
\newcommand{\PQ}{{\sss PQ}}
\newcommand{\QCD}{{\sss QCD}}
\newcommand{\eff}{{\rm eff}}

\newcommand{\mma}{\alpha}

\begin{document}

\title{Can QCD Axions Survive the Cosmological Constant Problem?}

\author[a]{Carsten van de Bruck,}

\author[b,c,d]{C.P.~Burgess}

\author[a]{and Adam Smith}

\affiliation[a]{School of Mathematical and Physical Sciences, University of Sheffield
}

\affiliation[b]{Perimeter Institute for Theoretical Physics
}
\affiliation[c]{Department of Physics \& Astronomy, McMaster University
}
\affiliation[d]{School of Theoretical Physics, Dublin Institute for Advanced Studies
}

\emailAdd{c.vandebruck@sheffield.ac.uk}
\emailAdd{cburgess@perimeterinstitute.ca}
\emailAdd{asmith69@sheffield.ac.uk}

\date{\Today}

\abstract{Mechanisms that dynamically relax the vacuum energy offer a concrete way to approach the cosmological constant problem, but because relaxation is not confined to the vacuum energy alone it can have consequences for the rest of low-energy physics. We explore this issue using the recently proposed `yoga' relaxation models as an explicit framework and show how relaxation differentially suppresses `slow' physics relative to a characteristic timescale set by the mass of the relaxon. It therefore need not alter {\it e.g.}~Higgs \& collider physics but can dramatically change how light scalar fields participate in cosmology. We revisit the QCD axion in this setting and show that the suppression of the axion's vacuum potential reshapes its behaviour on cosmological timescales while leaving fast, high–energy processes unaffected. The result is to alter the axion mass–coupling relation away from the standard QCD band, driving it into a regime already ruled out by observational constraints. In particular, suppression of the vacuum axion potential allows the QCD matter-induced potential to dominate even for matter densities relevant to cosmology and everyday matter, potentially driving the axion away from the CP-conserving minimum for QCD-motivated parameters. We conclude that conventional QCD axions are unlikely to remain viable in their standard form within vacuum-energy relaxation frameworks.}

\maketitle

\section{Introduction}

 \begin{quote}
 {\it Kein Operationsplan reicht mit einiger Sicherheit \"uber das erste Zusammentreffen mit der feindlichen Hauptmacht hinaus.} -- H. von Moltke
 \end{quote}
\begin{quotation}
{\it Everybody has plans until they get hit for the first time.} -- M. Tyson
 \end{quotation}

The cosmological constant problem is the elephant in the room of modern physics. Until recently the observed Dark Energy density seemed consistent with a Lorentz-invariant vacuum energy but nobody knew how to understand its small size in a technically natural way. This problem predated Dark Energy's discovery because the previous upper limits were already puzzlingly small compared with the fundamental scales of micro-physics (for reviews pre- and post-discovery see \cite{Weinberg:1988cp, Burgess:2013ara}). The puzzle only deepens if the evidence for Dark Energy being time-dependent continues to strengthen \cite{DESI:2024mwx}. The general attitude is pragmatic despair: everyone hopes the problem can be ignored -- perhaps dressed up with an anthropic perspective -- and works on something else, implicitly hoping that this other problem is not affected by whatever solves the cosmological constant problem. In this paper we argue that these hopes are likely to be dashed, particularly for those working in cosmology. 

The absence of a widely accepted solution to the cosmological constant problem is a hindrance to making this argument, so we here use a particular well-motivated framework to illustrate the more general point. To this end we re-examine the properties of the QCD axion \cite{Weinberg:1977ma, Wilczek:1977pj} -- as motivated as the low-energy residue of the symmetry-based Peccei-Quinn solution to the Strong-CP problem \cite{Peccei:1977hh} -- using as the illustrative framework the `natural relaxation' (or yoga) model \cite{Burgess:2021obw} that provides the low-energy effective description of a higher-dimensional mechanism \cite{Aghababaie:2003wz} for understanding the Dark Energy's small size (for a review see \cite{Burgess:2025vxs}). 

We show in particular how the usual inference of axion properties changes within this picture because the same physics that produces a small Dark Energy density also suppresses the axion's scalar potential, moving the axion mass and couplings into regions of parameter space that are already ruled out by existing constraints (see, for example, \cite{Arvanitaki:2010sy,Graham:2013gfa,Blum:2014vsa,Roussy:2020ily,Caputo:2024oqc,Baryakhtar:2025jwh}). In this scenario the QCD axion in particular is also ruled out as an explanation for Dark Matter \cite{Preskill:1982cy, Abbott:1982af, Dine:1982ah} (for a review see \cite{Marsh:2015xka}). 

How this conclusion is reached depends in part on precisely how the axion is embedded into the full theory. For the simplest embeddings of the Peccei-Quinn approach into the yoga framework the QCD axion line is moved to the left in the coupling-mass plane (compare Figs.~\ref{fig:fa vs mass} and \ref{fig:fa vs mass w yoga} below) and so falls into a well-constrained regime of parameter space. 

A similar conclusion arises for different reasons in dual versions of the PQ axion, providing a concrete illustration of how dual systems can in some ways be interestingly different \cite{Burgess:2025geh}. For the dual version the axion decay constant is Planck sized but the axion mass is larger than what would usually be expected for a QCD axion with this large a decay constant. Although this nominally puts it outside the reach of energy loss and direct detection experiments it instead falls prey to constraints that follow from the changed interplay between vacuum and matter potentials within dense objects that happens when the vacuum potential is suppressed. 

If the Dark-Energy motivated suppression of the scalar potential can change axion physics so dramatically one might worry that  the same arguments could also apply to other scalars, such as to the Higgs boson. We show why the same arguments do apply but why this is not a problem for Higgs physics. Suppression of the scalar potential is efficient for slow processes for which the relaxation field can adjust adiabatically, but inefficient for rapid processes where there is insufficient time for the relaxation field to respond. Consequently cosmology, being the standard testing ground for light scalars \cite{Olive:2001vz,Copeland:2006wr,Ferreira:2019xrr,Clifton:2011jh,Giare:2023kiv,DeAngelis:2023fdu,Baryakhtar:2024rky,Baryakhtar:2025uxs,Visinelli:2025gzg}, supplies the cleanest illustration of how and why relaxed axions can fail, independent of present--day searches and of complications inherent to particular astrophysical environments.

QCD itself generates both a vacuum and an in-matter potential for any axion motivated by solutions to QCD's strong-CP problem. In the usual telling of the story the in-matter potential is subdominant to the vacuum potential for matter densities smaller than nuclear densities, but this is no longer true within relaxation models for two reasons. First, once relaxation effects suppress the vacuum potential the in-matter potential can dominate for the baryon densities relevant to ordinary matter and to cosmology. Second, because the in-matter potential has a minimum that differs from the vacuum minimum one might expect the axion to take different values interior to matter than in the vacuum. But even for nuclear densities there is a trade-off between potential and gradient energies that implies a minimum size for objects in whose interiors this actually occurs.\footnote{In the standard story nuclei are too small for this to occur so phenomenologically interesting effects only occur for macroscopic objects like neutron stars  \cite{Hook:2017psm}.} This critical size also changes in relaxation models -- typically shrinking -- because they can change the energy cost of gradients.

We find relaxation of the axion's vacuum potential allows the axion field to be driven away from the CP-conserving minimum of the vacuum potential even inside ordinary objects throughout the post--BBN Universe. Such a displacement is incompatible with nuclear properties of stellar interiors and for early-Universe consistency. Within cosmology BBN and other primordial eras are sensitive to departures of QCD properties from the CP-conserving minimum \cite{Lee:2020tmi}, and the resulting bounds are strong enough to exclude relaxed QCD axions that evade other constraints. Alternatively, although axion properties can be chosen to maintain the in-matter alignment with the vacuum minimum throughout cosmic history, this requires parameter choices that spoil the relations that characterise a Peccei-Quinn QCD axion.

The rest of the paper is organised as follows. In \S\ref{sec:RelaxedCosmology} we briefly review the relaxation framework with which we work and explain how it suppresses contributions to the scalar potential (including contributions from the vacuum energy). An accidental approximate scaling symmetry plays an important role and this section also summarizes the properties of the light scalar dilaton to which it points. When an axion is embedded into this framework, the suppression of its vacuum potential leads to a trough-shaped joint dilaton-axion potential whose bottom is largely parameterized by the axion direction. There are a variety of ways to embed the axion into these models, of which we distinguish representative benchmark examples ({\it e.g.}~brane-localised versus bulk/saxionic axions), summarizing the induced non-derivative axion-matter couplings relevant for macroscopic environments. \S\ref{Fast vs Slow relaxation}  clarifies the differing physics of relaxation in fast vs slow regimes that explains why dramatic changes to cosmology can coexist with particle-physics processes remaining effectively unrelaxed. \S\ref{ssec:axionsinmasses} turns to constraints coming from the non-derivative interactions with matter, including from late-time cosmology where the baryon density is a known function of time. We present our conclusions and discuss implications and future directions in \S\ref{sec:Conclusions}.

\section{A relaxed attitude towards cosmology}
\label{sec:RelaxedCosmology}

The purpose of this section is to introduce the relaxation framework we use as an illustrative case for how the QCD axion potential can be dynamically suppressed as a side effect of a relaxation mechanism aimed at suppressing the cosmological constant problem. 

\subsection{Natural relaxation overview} 
\label{ssec:NaturalRelaxation}

Before diving into the details it is useful to summarize with broad brushstrokes the main features of the relaxation approach to understanding the small size of the Dark Energy. 

\subsubsection{The UV perspective}

In the UV the central idea is extra-dimensional: the gravitational backreaction of a 4D space-filling brane located within a 6D bulk can curve the extra dimensions and not curve the dimensions visible as Dark Energy to 4D cosmologists \cite{Carroll:2003db, Aghababaie:2003wz}. If all Standard Model (SM) particles were localized on such a brane their vacuum energy would contribute to the brane tension and this might explain why the curvature produced is not seen by cosmologists as a Dark Energy density. 

Of course this observation is only useful if the higher dimensions do not themselves have a cosmological constant, because in general brane-localized cosmologists would be expected to see this even if the brane tension is invisible. Interestingly (unlike in 4D) in six dimensions and higher a cosmological constant can be forbidden by even a single supersymmetry, making supersymmetric extra dimensions the minimal framework within which these ideas might work.\footnote{This has a counterpart in 4D because in 4D multiple supersymmetries can forbid a cosmological constant.} Higher dimensional supersymmetry need not be inconsistent with the observed lack of supersymmetry in particle colliders if supersymmetry is badly broken on the 4D brane on which ordinary particles are localized (as is not difficult to achieve). The low-energy world suggested by this picture consists of a non-supersymmetric particle physics sector (the brane) coupled to a supersymmetric\footnote{Brane-bulk couplings allow supersymmetry breaking to trickle from brane to bulk with gravitational strength, leaving an approximately supersymmetric bulk with SUSY breaking at the Kaluza-Klein scale.} gravitationally coupled dark sector (the bulk) \cite{Burgess:2004yq, Burgess:2021juk}. 

Detailed calculations within the higher dimensional theory \cite{Burgess:2011va, Burgess:2008yx, Bayntun:2009im, Burgess:2005cg, Hoover:2005uf, Burgess:2007vi, Williams:2012au, Burgess:2012pc} verify that loop corrections due to nonsupersymmetric brane-localized particles can be large despite often (but not always \cite{Tolley:2005nu, Tolley:2006ht, vanNierop:2011di}) leaving a flat effective low-energy 4D geometry. A sufficent condition for the brane tension not to curve the observable 4D world turns out to be the {\it absence} of a brane coupling to a particular 6D field (the 6D dilaton) \cite{Aghababaie:2003ar, Burgess:2011rv, Gautason:2013zw},\footnote{Significantly, the {\it absence} of a coupling between brane fields and a bulk field (like the dilaton, as is required for 4D flat geometries) is completely stable against arbitrary loops involving only brane fields, and this is part of the reason why Standard Model loops do not in themselves create naturalness problems for these models.} whose generic existence in 6D comes from an accidental approximate scaling symmetry generic to essentially all higher-dimensional supergravities \cite{Burgess:2020qsc}. 

Potentially dangerous loops of heavy bulk degrees of freedom turn out also to be benign because they are suppressed by supersymmetry right down to the bulk supersymmetry-breaking scale (the Kaluza-Klein -- or KK -- scale) \cite{Burgess:2005cg, Hoover:2005uf}. The most dangerous corrections involve bulk loops inducing on-brane couplings, but even these can be under control \cite{Williams:2012au, Burgess:2012pc}. The KK scale can even be in the right ball park if the extra dimensions are as large as they are phenomenologically allowed to be, for which the KK scale $1/r$ can be somewhat smaller than eV scales;\footnote{It is sometimes claimed that stronger limits exist on the allowed size of two large dimensions but these stronger bounds rely on having KK modes decay into observed particles, which is a model-dependent property not universal to all 6D theories.} a scenario called supersymmetric large extra dimensions (SLED) \cite{Aghababaie:2003wz}. 

For the present purposes what is important is that the core of the extra dimensional mechanism is a relaxation process. Explicit solutions \cite{Gibbons:2003di, Burgess:2004dh, Burgess:2011mt, Burgess:2014qha} show that as one changes the value of the tension on the 4D branes the extra dimensions adjust to compensate, while keeping the 4D spatial curvature unchanged.

\subsubsection{The IR perspective}

The above properties turn out to be reproduced by the low-energy effective 4D theory relevant to cosmology in these theories. In the 4D theory the accidental scaling symmetry is carried by a 4D Goldstone field $\tau$, whose expectation value also brings the 4D theory the news about hierarchies related to the size of the extra dimensions: $\tau \sim (M_6 r)^2$ where $M_6$ is the higher-dimensional Planck scale and $r$ is the extra-dimensional radius. If $M_6 \sim 100$ TeV and $1/r \sim 1$ eV, as for models with two large extra dimensions, then $\tau \sim 10^{28}$. 

Dimensional reduction ensures $r$ also determines the size of the 4D Planck scale $M_p \sim M_6^2 r$ and so when the 4D theory is written in 4D Einstein frame ordinary particles couple to $\tau$ only through a Jordan-frame (brane) metric $\tilde g_{\mu\nu} \propto g_{\mu\nu}/(M_6 r)^2 = g_{\mu\nu}/\tau$, ensuring that $\tau$ couples to ordinary (brane-localized) matter like a Brans-Dicke scalar. This implies in particular that the Higgs expectation value -- and so also all SM particle masses -- share a common $\tau$-dependence when measured in units of the Planck mass, $M_\EW \sim M_p/\tau^{1/2} \sim M_6$, up to dimensionless coupling factors.\footnote{This in itself is not quite sufficient to ensure all SM mass ratios are $\tau$-independent because the dominant contribution to hadron masses comes from the QCD scale $\Lambda_\QCD$. But $\Lambda_\QCD \propto e^{-4\pi/\alpha(M)} M$ is itself generically proportional to a UV mass scale, and so also scales like $1/\tau^{1/2}$ if this UV mass scale does (as is easy to arrange).} 

If neutrinos receive masses from the Weinberg dimension-5 SMEFT operator localized on a brane (or through a see-saw type mechanism, both of which are quadratic in the Higgs vev) then their masses are order $m_\nu \sim M_p/\tau \sim M_6/\tau^{1/2} \sim 1/r$. All of these mass scales make sense (in order of magnitude) for $\tau \sim 10^{28}$ (which is not really a surprise because they encode in 4D what two large dimensions are known to do in 6D).  

Extra-dimensional relaxation can also be captured within the low-energy 4D effective theory by integrating in a representative modulus field, $\phi$, (the relaxon), whose evolution captures how the extra-dimensional degrees of freedom adjust to seek a vanishing 4D effective vacuum energy density.\footnote{In two-brane models a natural candidate for $\phi$ is the distance between the two branes, and the resulting 4D theory is a special case of the Yoga models proposed in \cite{Burgess:2021obw} independent of the above 6D motivations.} The large size of $\tau$ implies the 4D scalar potential for the dilaton--relaxon system can be usefully organized as an expansion in inverse powers of $\tau$, which turns out to have the form \cite{Burgess:2021obw}  
\begin{equation}\label{Expintau}
V(\phi,\tau) = \frac{M_p^4}{\tau^2} \left[ \left| w_x - \frac{c_1 w}{\tau} \right|^2 + \frac{c_2 |w|^2}{\tau^2} \;+\; \cO(\tau^{-3}) \right] \,,
\end{equation}
where the $c_i$ are order-unity dimensionless functions of $\phi$ (with possibly a weak logarithmic dependence on $\tau$). These terms have their origins in the K\"ahler potential describing the low-energy limit of the supersymmetric bulk, while $w_x = W_x/M_p^2$ and $w = W_0/M_p^3$ are pieces of its superpotential (in Planck units).\footnote{Expressions for $c_1$ and $c_2$ are given in \cite{Burgess:2021obw}, though the result for $c_2$ is misstated (and corrected in \cite{Cicoli:2024bwq}). K\"ahler functions and superpotentials appear even though the particle-physics sector is not supersymmetric at all because couplings between the bulk and particle-physics sectors are described using the formalism of \cite{Komargodski:2009rz}.}

Comparison with the higher-dimensional construction shows that the $|w_x|^2$ term describes the brane tension contribution. The perfect-square structure of the first term in the square brackets in \pref{Expintau} is a consequence of the approximate supersymmetry of the gravitationally coupled bulk sector and its vanishing corresponds in the higher-dimensional theory to a BPS-like choice of brane-bulk couplings for which the brane does not break the bulk supersymmetry \cite{Williams:2012au, Burgess:2012pc}. Notice that in order of magnitude the leading term is $M_p^4/\tau^2 \sim M_\EW^4 \sim M_6^4$ when the dimensionless quantity $w_x$ is order unity, and this is the right size for a brane tension (and way too large to describe the Dark Energy). The perfect-square form guarantees it to be positive. 

The key point is that extra-dimensional relaxation corresponds in this theory to minimizing $V(\phi, \tau)$ with respect to $\phi$ (or to all similar moduli if there is more than one). For $\tau \sim 10^{28}$ the $1/\tau$ expansion is {\it really} good and if $1/\tau^4$ terms can be neglected varying $\phi$ very generally sets the squared term to zero. When $1/\tau^4$ terms cannot be neglected (such as when identifying the size of $V$ at the minimum) the $1/\tau^4$ contribution is the leading effect. This is how the low-energy theory sees that extra-dimensional back-reaction drives the 4D curvature to zero (or to be very small once $1/\tau^4$ terms are included) as underlying properties like the brane tension ({\it i.e.}~$W_x$) are varied.

Supersymmetry aficionados might worry that allowing $\phi$ to drive $W_x$ to near zero amounts to dynamically restoring supersymmetry, since the things that get squared in supersymmetric scalar potentials tend to be the auxiliary fields that parameterize how badly supersymmetry is broken. Any argument that explains a small scalar potential by making the world very supersymmetric is not good enough to describe the world around us. It is here that gravity plays a key role, because in supergravity it is really $D_x W =  W_x + K_x W/M_p^2$ that quantifies supersymmetry breaking, and much of the point of the complications of ref.~\cite{Burgess:2021obw} is to arrange $D_xW$ to be large while having $W_x$ small.  

There is also a nice story \cite{Burgess:2021obw, Burgess:2022nbx} as to why the $c_i$'s can contain polynomials of $\log\tau$ and how this dependence can generate a minimum for $\tau$ at $\tau \sim 10^{28}$ without requiring ridiculous choices for parameters in the potential. For instance if they are quadratic functions of $\log \tau$ then the potential becomes
\begin{equation}\label{axio-dil pot2}
    V(\tau,\mfa) = \frac{M_p^4}{\tau^4} \Bigl( v_0 -v_1 \log \tau + \frac{v_{2}}{2}\log^2\tau + \cdots \Bigr)  \,,
\end{equation}
for some coefficients $v_i$. For $v_i \sim \cO(50)$ the resulting potential can easily have a minimum with $\tau_m \sim 10^{28}$ and $V(\tau_m)$ positive as required for successful phenomenology. 

Better yet, the value of the potential at this minimum is naturally of order $M_p^4/\tau^4 \sim (M_\EW^2/M_p)^4 \sim 1/r^4$ which scales with $\tau$ as does the Dark Energy, leading to a plausibly small result when $\tau \sim 10^{28}$ ({\it i.e.}~when the extra dimensions are large). Large $\tau$ also ensures that the mass of the $\tau$ field, $m_\tau \sim M_p/\tau^2$, is of order the present-day Hubble scale, predicting a time-dependent Dark Energy of the Albrecht-Skordis type \cite{Albrecht:1999rm, Albrecht:2001xt}.

Intriguingly, cosmological analysis of a minimal version of this model \cite{Smith:2024ibv} (with a non-QCD axion playing the role of Dark Matter) provide a better fit to current datasets than does $\Lambda$CDM  \cite{Smith:2025uaq}, largely because the evolution of $\tau$ with time makes particle masses differ at recombination and this turns out to reduce the Hubble tension. Although not proposed with this in mind, the model turns out to provide a dynamical realization of the general mechanism proposed in \cite{Sekiguchi:2020teg}.

This is not to say that these models have no flaws, the main one of which seems to be that the Brans-Dicke-like force mediated by the $\tau$ field is too strong to be consistent with solar-system tests of General Relativity. Work is in progress to see whether this can be fixed, such as through screening mechanisms along the lines of \cite{Brax:2023qyp}.

\subsection{Relevance to the PQ axion}
\label{ssec:PQRelevance}

The natural relaxation story is all very nice, but what does it have to do with the QCD axion? The point is that the relaxation model described above tends to suppress the entire low-energy scalar potential rather than just the cosmological constant. In the presence of an axion field $\mfa$ the same arguments go through but now in expressions like \pref{Expintau} the coefficients $c_i = c_i(\phi,\mfa)$ and $w_x = w_x(\phi,\mfa)$ {\it etc} are functions of both the relaxation modulus {\it and} the axion. The minimization of $V$ with respect to $\phi$ then produces an axion-dependent solution $\phi = \phi_m(\mfa)$, rather than a specific value. 

Once evaluated at this solution the potential is once again suppressed by at least $1/\tau^4$ for all values of the axion field. Effectively the scalar potential has a trough-like shape with $\phi_m(\mfa)$ parameterizing its minimum. The potential along the bottom of this trough is not exactly zero, but is supressed by the same power of $1/\tau$ that suppresses the Dark Energy density. As a consequence the direction along the bottom of the trough ($\mfa$ in the above example) is much shallower than would have been naively expected, leading to a suppression of the axion mass relative to naive expectations. We return in \S\ref{Fast vs Slow relaxation} to the question of whether the same arguments also apply to other scalars (they do) and why this nevertheless need not undermine our understanding of {\it e.g.}~Higgs physics in colliders.  

\subsubsection{Vanilla QCD axion}

In the usual telling of the tale the QCD axion is a Goldstone boson for a global $U(1)$ Peccei-Quinn symmetry that has a QCD anomaly. The strong couplings of the QCD vacuum then conspire to generate a low-energy effective lagrangian for the axion of the general form 
\be \label{vanillaAxion}
   - \frac{\cL_{\rm ax}}{\sqrt{-g}} = \frac12 \, f^2 \, g^{\mu\nu} \partial_\mu \mfa \, \partial_\nu \mfa + V_\PQ(\mfa) - \partial_\mu \mfa \, J^\mu + \cdots \,,
\ee
where $f$ is a characteristic scale at which the PQ symmetry is spontaneously broken, $J^\mu$ is a dimension-3 current built from ordinary matter fields (possibly including electromagnetic Chern-Simons terms) and the scalar potential is
\be \label{PQV}
 V_\PQ(\mfa) = \Lambda_\QCD^4 \, U(\mfa) \,,
\ee
where $\Lambda_\QCD \simeq 200$ MeV is the characteristic scale of QCD and $U(\mfa)$ is a periodic order-unity function of the dimensionless axion field $\mfa$ whose detailed form as given explicitly in \cite{GrillidiCortona:2015jxo} is
\be \label{QCDVacuumPotential}
  U_{\rm vac}(\mfa) =- \sqrt{1 - \frac{4 m_u m_d}{(m_u + m_d)^2} \, \sin^2 \left(\frac{\mfa}{2} \right) } \,,
\ee
where $m_u$ and $m_d$ are the up and down quark masses. This approaches $U_{\rm vac} \simeq -\left| \cos \left( \frac{\mfa}{2} \right) \right|$ in the limit when $m_d \simeq m_u$.  

\begin{figure}[hbtp!]
    \centering
    \includegraphics[width=0.9\linewidth]{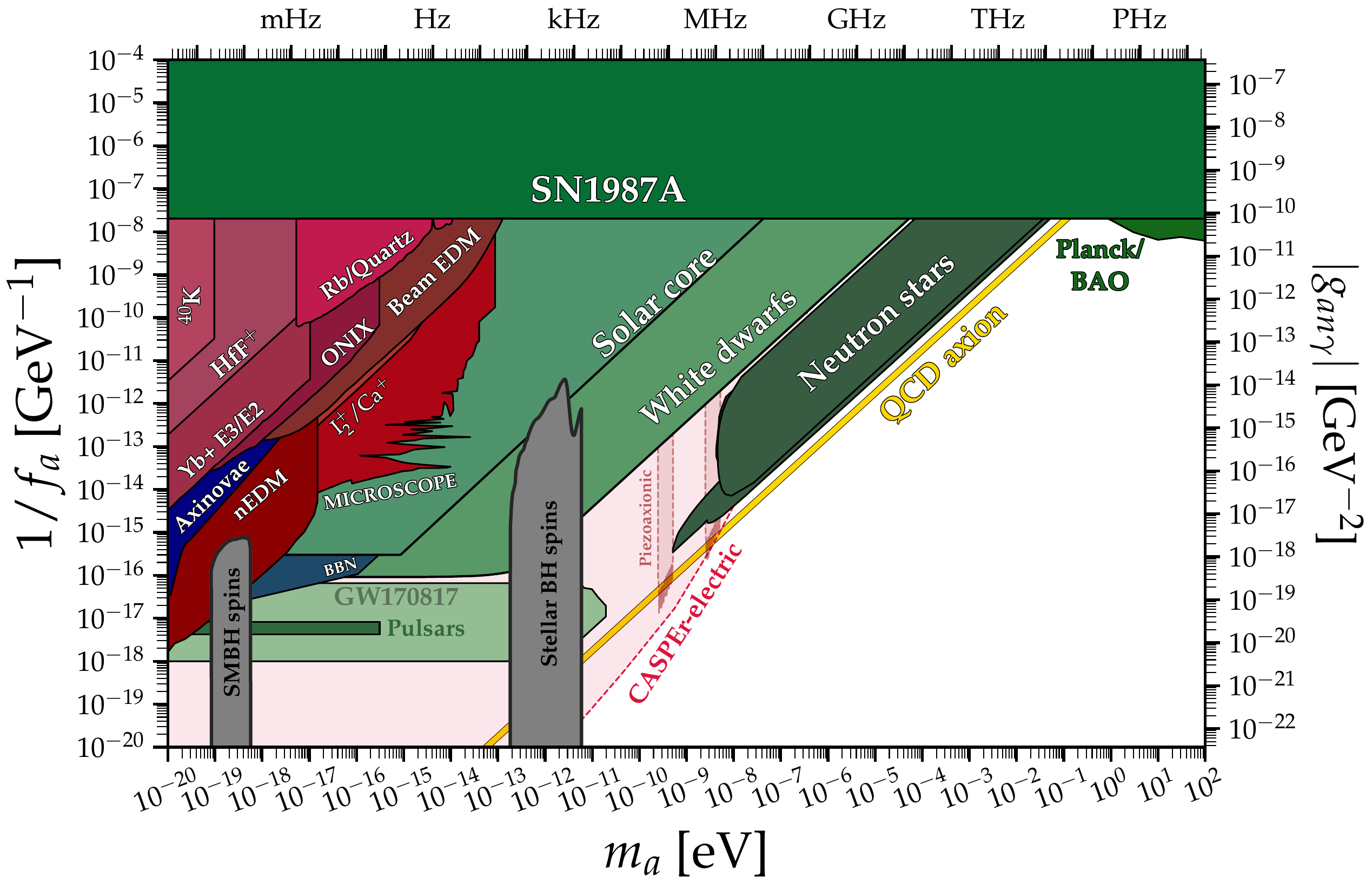}
    \caption{Parameter space constraints for the QCD axion taken from \cite{AxionLimits}, with the canonical QCD band defined by the relations \pref{QCDmavsfa} and \pref{VanillaAxCoupling} shown in yellow.   }
    \label{fig:fa vs mass}
\end{figure}

Canonically normalizing by redefining $\mfa = \mma/f$ and assuming the second derivative $U''$ to be order unity at its minimum gives the usual axion mass estimate
\be \label{QCDmavsfa}
   m_a \simeq \frac{\Lambda_\QCD^2}{f},
\ee
and shows that axion matter couplings have strength 
\be \label{VanillaAxCoupling}
  \cL_{\rm int} = \frac{1}{F_a} \, \partial_\mu \mma \, J^\mu \qquad \hbox{with} \qquad F_a = f\,.
\ee
For the specific case\footnote{For traditional axion models choosing the decay constant much larger than $10^{9}$ GeV would normally be disfavoured by axion dynamics during and after inflation (for a review see \cite{Marsh:2015xka}). But in the UV theory considered here the nature of cosmology at inflationary scales is likely to be very different from the vanilla picture usually used -- see {\it e.g.}~\cite{vanNierop:2011di, Burgess:2022nbx, Cicoli:2024bwq} for preliminary discussions -- requiring these arguments to be reassessed.} $F_a = f = M_p$ eq.~\pref{QCDmavsfa} predicts $m_a \sim 10^{-11}$ eV.  Exclusion plots for various choices of $F_a$ (that are only really precise once specific choices for $J^\mu$ are made and any numerical factors in the current $J^\mu$ are specified) and $m_a$ are summarized in plots like Fig.~\ref{fig:fa vs mass}, on which the relation \pref{QCDmavsfa} appears as a diagonal yellow line. 

\subsubsection{Relaxation of the potential}

However if the physics describing PQ symmetry breaking is also localized on the brane (in the UV theory) then the axion enters into the low-energy theory with the same Jordan-frame metric as does everyone else, ensuring that it enters the potential as a contribution to $w_x$ (and possibly to the $c_i$'s). 

For instance, suppose in the absence of an axion $w_x$ passes through zero for $\phi = u$. Then -- keeping in mind that $W_x$ has dimension (mass)${}^2$ -- near this zero it can be written $W_x \simeq m (\phi - u)$. Suppose to this is added a periodic dependence on the axion field,\footnote{See \cite{Komargodski:2009rz} for the rules on what types of fields quantities like $W_x$ can depend on. The coefficient of the axion term defines the scale $\Lambda$. It can be chosen to be positive by changing the sign of $\mfa$.} like  
\be \label{WxAxion}
  W_x \simeq m ( \phi - u) + \Lambda^2 \sin\left( \frac{\mfa}{2} \right)\,,
\ee
then the leading tension term in \pref{Expintau} becomes
\be \label{tausq}
   \frac{|W_x|^2}{\tau^2} = \frac{1}{\tau^2} \left[ m^2 |\phi - u|^2 + \tfrac12 \Lambda^4 \Bigl(1 - \cos \mfa \Bigr) + 2m \Lambda^2 \hbox{Re} ( \phi  - u)  \sin \left( \frac{\mfa}{2} \right) \right] \,.
\ee
If the dimensional factors are $m \sim M_p$ and $\Lambda \sim \varepsilon M_p$ then the first term expanded about $\phi = u$ gives a mass to $\phi$ of size $m_\phi \sim M_p/\tau \sim 1/r$, assuming the $\phi$ kinetic terms are $\tau$-independent (which is true in 4D Einstein frame for a mode coming from the bulk sector of the UV theory). This size a mass is of order the KK scale, as might be expected for a bulk mode. The second term gives an axion potential of the form \pref{PQV}, with a commonly used choice for $U(\mfa)$ (though not the one predicted by QCD \cite{GrillidiCortona:2015jxo}) that corresponds to an effective QCD scale $\Lambda_{\rm eff} \simeq \Lambda/\sqrt\tau \sim \varepsilon M_p/\sqrt\tau \sim \varepsilon M_\EW$ (suggesting $\varepsilon \sim 10^{-5}$ if $\Lambda_{\rm eff}$ is to be the QCD scale  ($\sim 0.1$ GeV) and $M_\EW \sim 10$ TeV). 

Perhaps more importantly, adding a constant to $W_x$ just changes the value $u$ taken by $\phi$ at the vanishing point of $W_x$ but does not remove the property that the potential vanishes at its minimum (at least at leading order in $1/\tau$). This means that (at order $1/\tau^2$)  the effective potential given by evaluating $W_x[\phi(\mfa)]$ at the minimum for $\phi$ is completely independent of $\mfa$, implying the axion would be massless (although its non-derivative interactions need not in general vanish). This masslessness does not survive once $1/\tau^4$ terms are included, but the result can be very different from what would naively be expected from \pref{tausq}.

\begin{figure}[htbp]
    \centering
    \includegraphics[width=\linewidth]{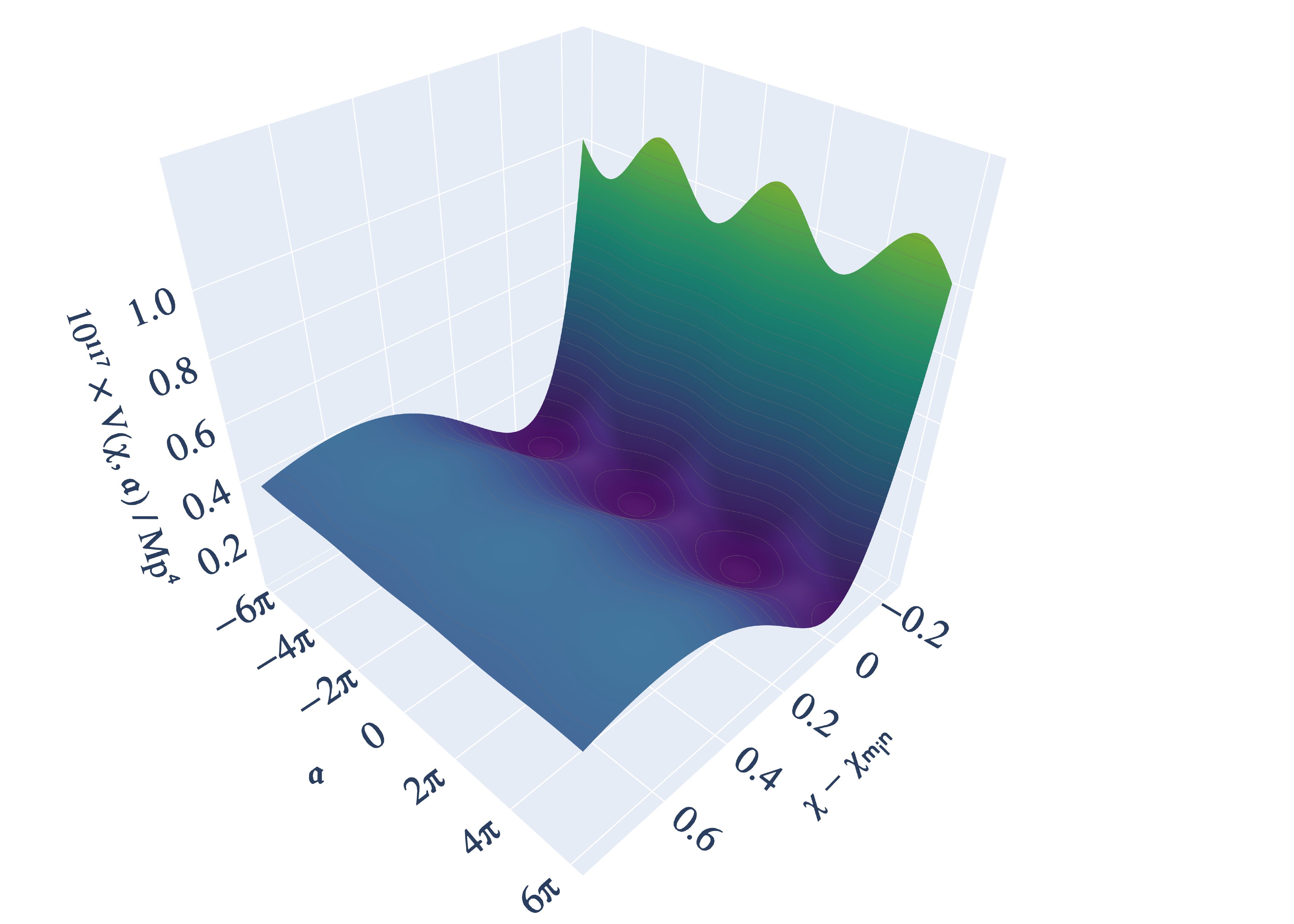}
    \caption{Surface plot of the axion--dilaton potential (\ref{axio-dil pot0}) in the $(\chi = \sqrt{\frac{3}{2}}\ln(\tau)$, $\mfa$) plane for
    $\varepsilon = 10^{-3}$, and the same values for the constants $v_i$ as were used in \cite{Burgess:2021obw}. A trough is clearly visible stabilising the dilaton direction, while the axion direction exhibits periodic maxima and minima.}
    \label{fig:trough}
\end{figure}

At order $1/\tau^4$ the potential in general acquires an axion dependence even after $\phi$ has been minimized, assuming that terms like $c_2 |w|^2$ in \pref{Expintau} also depend on $\phi$. In particular, if the $\phi$-derivatives of $c_2 |w|^2$ are not small at the value of $\phi$ that makes the first perfect square in \pref{Expintau} vanish then the leading dependence on $\mfa$ in $V[\phi_m(\mfa), \mfa]$ replaces \pref{axio-dil pot2} with
\begin{equation}\label{axio-dil pot0}
    V[\phi_m(\mfa),\mfa] =  \frac{M_p^4}{\tau^4}  \left[ v_0 -v_1 \log \tau + \frac{v_{2}}{2}\log^2\tau 
    + \varepsilon^2 v_3 \sin \left(\frac{\mfa}{2}\right) + \cdots \right]  \,,
\end{equation}
where again $v_i$ are dimensionless constants not larger than order 50. This resulting shape of the potential is shown in figure \ref{fig:trough}, which shows that the dilaton local minimum in the trough remains intact for all values of $\mfa$ even when $\varepsilon = 10^{-3}$. The new axion term using QCD-motivated values does not destabilise the minimum for the dilaton.

What is important in this expression is the size of the overall coefficient and not the specific periodic function of $\mfa$ that arises (which depends on the initial choice made for $W_x$). Comparing the axion-dependent part with \pref{PQV} shows that the effective scale of the potential has been reduced from the unrelaxed value $\Lambda_{\rm unr}^4 \sim \Lambda^4/\tau^2 \sim \varepsilon^4 M_\EW^4 = \Lambda_\eff^4$ seen in \pref{tausq} to the much smaller relaxed value $\Lambda_{\rm rel}^4 \sim \Lambda^2 M_p^2/\tau^4 \sim \varepsilon^2 M_\EW^4/\tau^2 \sim \varepsilon^2/r^4$. Recall that the present-day Dark Energy density in these models is of order $M_p^4/\tau^4 \sim 1/r^4$ and so the scale appearing in the relaxed axion potential is suppressed relative even to this by powers of $\varepsilon$. 

\subsection{Benchmark axion embeddings}

The size of the axion mass and decay constant corresponding to the suppressed potential depend on the $\tau$-dependence of the axion kinetic term, and there are several reasonable choices that can be made for this. This section outlines two representative cases in particular.

\subsubsection{Brane axion}

The first benchmark case we consider applies when the axion in question is a scalar PQ field localized on a 4D brane (as are all Standard Model particles). This is the natural way a PQ axion would arise within the above relaxation framework. In this case the axion lagrangian is dilaton-independent in the same Jordan frame as for any Standard Model particles (like the Higgs), since this corresponds to the higher-dimensional Einstein frame. Since the leading term in the potential \pref{tausq} is proportional to $1/\tau^2$ the metric rescaling required to make it $\tau$-independent is 
\be \label{JordanFrameDef}
    g_{\mu\nu} \propto \tau \tilde g_{\mu\nu} \,.
\ee
The requirement that $\tau$ is also removed from the axion kinetic term when expressed in terms of $\tilde g_{\mu\nu}$ implies that in 4D Einstein frame it must scale with $\tau$ as
\be
 - \frac{ \cL_{\rm kin}}{\sqrt{-g}}  \sim \frac{f_b^2}{\tau} \, g^{\mu\nu} \partial_\mu \mfa \, \partial_\nu \mfa \qquad \hbox{(brane-localized axion)} \,,
\ee
for some mass scale $f_b$ and so $f= f_b/\sqrt\tau$. 

As a check on this we compute the axion mass in the absence of relaxation -- {\it i.e.}~using the potential \pref{tausq} -- for which $\Lambda_{\rm unr}^2 \sim \Lambda^2/\tau = \varepsilon^2 M_p^2/\tau \sim \varepsilon^2 M_\EW^2$ (which agrees with $\Lambda_\QCD^2$ when $\varepsilon \sim 10^{-5}$). With this choice the unrelaxed axion mass becomes $\hat m_a 
\sim \varepsilon^2 M_p^2/(f_b\sqrt\tau)$, so $\Lambda_{\rm unr}$, $f$ and $\hat m_a$ scale with $\tau$ in precisely the same way as do all other Standard Model masses, ensuring that their ratios are $\tau$-independent. 

Including relaxation lowers the above mass prediction to $m_a \sim \Lambda_{\rm rel}^2/f$ so using $\Lambda_{\rm rel}^2 \sim \varepsilon M_p^2/\tau^2$ implies the relaxed axion mass $m_a \sim \varepsilon M_p^2/(f_b \tau^{3/2})$ scales differently with $\tau$ than do the masses of Standard Model particles. Numerically, if $f_b \sim M_p$  we find $f \sim f_b/\sqrt\tau \sim M_\EW$ and $m_a \sim  \varepsilon M_p/\tau^{3/2} \sim \varepsilon M_\EW/\tau$, which is order $10^{-20}$ eV for the QCD case (for which $\varepsilon \sim 10^{-5}$) when $\tau \sim 10^{28}$. 

The coupling $F_a$ can be computed in a similar way, assuming that the axion-SM coupling is also $\tau$-independent in the SM Jordan frame. To see what this implies we write down a few representative terms in a 4D brane-localized lagrangian, such as
\bea
   - \frac{\cL_{\rm b}}{\sqrt{-\tilde g}} &=&  \tfrac12 f_b^2 \, \tilde g^{\mu\nu}  \partial_\mu \mfa \, \partial_\nu \mfa + \tfrac14 \tilde g^{\mu\lambda} \tilde g^{\nu\rho} F_{\mu\nu} F_{\lambda\rho}+{\tilde{e}^\mu{}}_a \ol \psi \gamma^a D_\mu \psi + m_0 \, \ol \psi \psi \nn\\
   &&\qquad\qquad\qquad + c_1 {\tilde{e}^\mu{}}_a \partial_\mu \mfa \; \ol \psi \gamma^a \psi + c_2 \mfa \; \tilde \epsilon^{\mu\nu\lambda\rho}  F_{\mu\nu}F_{\lambda\rho} + \cdots \,,
\eea
where $c_i$ denote dimensionless order-unity constants and the inverse vierbein satisfies ${\tilde{e}^\mu{}}_a{\tilde{e}^\nu{}}_b \eta^{ab} = \tilde g^{\mu\nu}$ and ${\tilde{e}^\mu{}}_a {\tilde{e}^\nu{}}_b \tilde g_{\mu\nu} = \eta_{ab}$ while $\psi$ and $F_{\mu\nu} = \partial_\mu A_\nu - \partial_\nu A_\mu$ are representative brane-localized spin-half and spin-one fields. Switching to Einstein frame using $g_{\mu\nu} = \tau \tilde g_{\mu\nu}$ then gives
\bea\label{stringframebraneaction}
   - \frac{\cL_{\rm b}}{\sqrt{-g}} &=&  \frac{f_b^2}{2\tau} \,   g^{\mu\nu}  \partial_\mu \mfa \, \partial_\nu \mfa + \tfrac14   g^{\mu\lambda}   g^{\nu\rho} F_{\mu\nu} F_{\lambda\rho} + \frac{1}{\tau^{3/2}}{ {e}^\mu{}}_a \ol \psi \gamma^a D_\mu \psi +\frac{ m_0}{\tau^2} \, \ol \psi \psi \nn\\
   &&\qquad\qquad\qquad + \frac{ c_1}{\tau^{3/2}} { {e}^\mu{}}_a \partial_\mu \mfa \; \ol \psi \gamma^a \psi + c_2 \mfa \;   \epsilon^{\mu\nu\lambda\rho}  F_{\mu\nu}F_{\lambda\rho} + \cdots \,,
\eea
confirming $f = f_b/\sqrt\tau$ as before. Canonically normalizing shows that fermion masses are $m_0/\sqrt\tau$ as expected, while comparing with \pref{VanillaAxCoupling} shows that the decay constant, $F_a$, appearing in the axion-fermion derivative coupling and the decay constant, $F_\gamma$, appearing in the axion-photon couplings are
\be \label{braneFa}
   F_a \sim F_\gamma \sim  f \sim \frac{f_b}{\sqrt\tau} \,.
\ee
As mentioned earlier, this $\tau$-dependence ensures the ratio of ordinary particle masses to $F_a$, $F_\gamma$ and the {\it unrelaxed} axion potential scale $\Lambda_{\rm unr}$ and unrelaxed axion mass $\hat m_a$ are $\tau$-independent, but the same is {\it not} true for their ratio to the relaxed scale $\Lambda_{\rm rel}$ or the relaxed axion mass, $m_a$, since these depend differently on $\tau$.

\begin{figure}[hbtp!]
    \centering
    \includegraphics[width=0.9\linewidth]{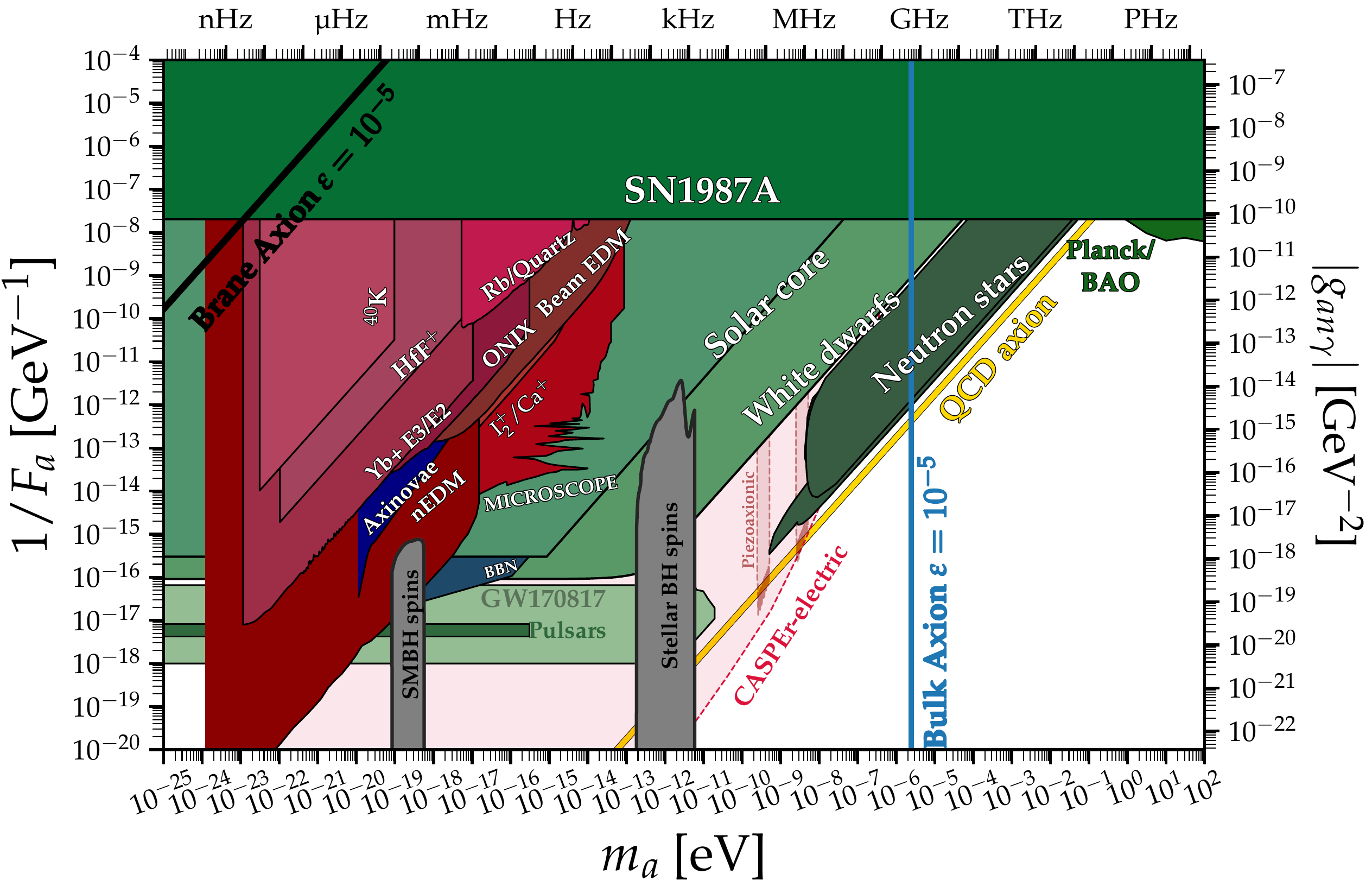}
    \caption{An overlay of the constraints from Fig.~\ref{fig:fa vs mass} with the modified QCD axion superimposed for both the bulk and brane axion variants of the yoga relaxation mechanism shown in blue and black respectively. }
    \label{fig:fa vs mass w yoga}
\end{figure}

Numerically \pref{braneFa} predicts $F_a \sim M_\EW$ when $f_b \sim M_p$ and the product $F_a m_a$ is independent of the free parameter $f_b$ and so the straight line swept out in the $1/F_a$-$m_a$ plane as $f_b$ is varied is parallel (in a log-log plot) to the traditional QCD axion line  (as sketched in Fig.~\ref{fig:fa vs mass w yoga}). The suppression of the axion mass moves the QCD axion line into a region of the $1/F_a$-$m_a$ plane that is well constrained by particle phenomenology.

\subsubsection{Bulk axion}

Another natural choice for the $\tau$-dependence of the axion kinetic term is whatever would be expected in the UV theory  when the axion is the partner of $\tau$ under bulk supersymmetry (sometimes called a {\it saxion}), along the lines studied in \cite{Smith:2024ibv, Smith:2024ayu, Smith:2025grk}. In this case bulk supersymmetry implies the kinetic term is
\be\label{bulk_ax_kin}
 - \frac{ \cL_{\rm kin}}{\sqrt{-g}}  \sim \frac{M_p^2}{\tau^2} \, g^{\mu\nu} \partial_\mu \mfa \, \partial_\nu \mfa \qquad \hbox{(saxion)} \,,
\ee
which when compared with \pref{vanillaAxion} gives 
\be \label{fmsizebulk}
  f \sim \frac{M_p}{\tau} \sim \frac{1}{r} \qquad \hbox{and so} \qquad m_a \sim \frac{\Lambda_{\rm rel}^2}{f} 
\sim \frac{\varepsilon M_p}{\tau} \sim \frac{\varepsilon}{r} \,. 
\ee
This type of axion has a mass scale that (for $\varepsilon \leq 1$) is at most of order the Kaluza-Klein (or Dark Energy) scale, as might have been expected for a bulk field. The QCD choice $\varepsilon \sim 10^{-5}$ and taking $1/r$ of order 0.1 eV (for large-dimensional models) then implies $m_a \sim 10^{-6}$ eV.

The prediction for the decay constant $F_a$ appearing in matter couplings is made in a similar way. It turns out to be important when doing so that this particular axion starts its life in the UV theory as a 2-form gauge potential $B_{\mu\nu}$, that is dual\footnote{Dualization for axions with a scalar potential is described in \cite{Quevedo:1996uu, Dvali:2005an} and some aspects of how to dualize the QCD axion in particular are discussed in \cite{Dvali:2022fdv, Burgess:2023ifd}.} to a scalar in 4D. To compute $F_a$ take the simplest coupling between a 2-form field and brane-localized matter as 
\be
  \cL_{\rm int} = \frac{1}{f_\ssB^2} \, G \wedge \tilde J \,, 
\ee
where $G = \exd B$ is the 2-form field strength (canonically normalized in 6D) and $\tilde J^\mu$ is a brane-localized Standard Model current, such as the fermion bilinear\footnote{In this case gauge invariance forbids the current from being a brane-localized electromagnetic Chern-Simons term, though an electromagnetic anomaly in this language can be encorporated as a term $B\wedge F$.} 
\be\label{tildeJdef}
   \tilde J_\mu = {{\tilde{e}}^a}_\mu \, \ol\psi \gamma_a \, \psi = \frac{1}{\sqrt \tau}\, {e^a}_\mu \, \ol\psi \gamma_a \, \psi \, 
\ee
where $\psi$ is a brane-localized spin-half fermion and the last equality converts to 4D Einstein frame as in the discussion surrounding \pref{stringframebraneaction}. This lagrangian is to be integrated over the 4D world-surface of the brane, so assuming $\tilde J^\mu$ has dimension (mass)${}^3$ -- as would a fermion bilinear -- requires the coupling for this interaction to have dimension (mass)${}^{-2}$.  $f_\ssB$ denotes the appropriate UV scale that appears in this way, whose natural upper limit in the Einstein frame would be the 4D Planck scale $M_p$ (corresponding to a Jordan-frame value being the 6D Planck scale $M_6$). 

Performing the duality then leads to an axion-matter coupling of the form \cite{Burgess:2021obw, Brax:2022vlf, Burgess:2025geh} 
\be
   \cL_{\rm ax} = - \tfrac12 \sqrt{-g} \; f^2 \, \left( \partial_\mu \mfa + \frac{\tilde J_\mu}{f_\ssB^2} \right) \left( \partial^\mu \mfa + \frac{\tilde J^\mu}{f_\ssB^2} \right) \,,
\ee
with $f$ as given in \pref{fmsizebulk}. For the case of a fermion bilinear current, comparing the above to the fermion kinetic term $\tau^{-3/2} \ol \psi \gamma^\mu \partial_\mu \psi$ and keeping in mind \pref{tildeJdef} reveals that canonically normalizing $\mfa$ and $\psi$ and comparing to \pref{VanillaAxCoupling} then implies
\be
   F_a \sim \frac{f_\ssB^2}{\tau f} \sim \frac{f_\ssB^2}{M_p}   \,,
\ee
which becomes $F_a \sim M_p$ when $f_\ssB \sim M_p$ as might have been expected for a gravitationally coupled KK mode. Having $F_a$ be proportional to $1/f$ rather than $f$ is characteristic of duality transformations and makes $F_a$ large despite $f$ as predicted in \pref{fmsizebulk} being much smaller than normally considered in axion physics.

Two things are noteworthy about these expressions. First, notice that the axion mass when $F_a \sim M_p$ is $10^{-6}$ eV -- see the discussion below eq.~\pref{fmsizebulk} -- and so despite having $\Lambda_{\rm rel} \ll \Lambda_{\rm unr}$ the axion mass is {\it larger} than the prediction \pref{QCDmavsfa} for the vanilla QCD axion. From a UV point of view this happens because having the axion kinetic term arise in the bulk changes the relation between $f$ and $F_a$ relative to the brane case. Because the scale $f_\ssB$ does not appear at all in the expression for $m_a$ in this case $f_\ssB$ does not cancel in the product $F_a m_a$. Consequently the curve traced in the $1/F_a$ vs $m_a$ plane as $f_\ssB$ is varied is vertical and not parallel to the standard QCD axion line (see Fig.~\ref{fig:fa vs mass w yoga}).  

It is also noteworthy that the phenomenological constraints shown in the figure coming from derivative interactions like \pref{VanillaAxCoupling} largely carry no teeth to the right of the usual QCD axion line, such as when $F_a \sim M_p$ and $m_a \sim 10^{-6}$ eV. This potentially makes it more interesting to explore such an axion's late-time cosmology and any non-derivative couplings with ordinary matter. Indeed, QCD provides a concrete example for which these non-derivative couplings can be explicitly computed \cite{Cohen:1991nk}, a topic to which we return in \S\ref{ssec:axionsinmasses} below after first being more precise about which kinds of masses get suppressed by the above relaxation mechanism.

\section{Relaxation, Fast and Slow}
\label{Fast vs Slow relaxation}

The general suppression of the entire scalar potential is a common objection to relaxation mechanisms aimed at the cosmological constant problem. How does one suppress the Dark Energy density without removing all other successful predictions obtained using scalar potentials, such as for Higgs physics or inflationary cosmology? This section sketches out the main underlying reason why this need not be a show-stopper.  

The main thought is that the efficiency of relaxation in any particular physical process depends on the time-scale over which that process takes place. Relaxation is in essence a dynamical thing and so must respond as fields change with time. The key question is whether or not the timescale of interest is shorter or longer than the characteristic relaxation time, $t_r$. 

For fast processes -- like Higgs scattering or decays --   that take much less than $t_r$ to complete one can work in the `sudden' approximation: the process completes so quickly that the relaxation field has no time to respond. The scalar potential relevant to this case is the naive (unsuppressed) one. But for slow processes -- like for axion cosmology -- that evolve over times much longer than $t_r$ the adiabatic approximation is most appropriate: the relaxon field has lots of time to track the motion of any other fields in the problem and so can efficiently suppress the scalar potential throughout.\footnote{This distinction between slow adiabatic adjustment and fast evolution that directly excites heavy degrees of freedom is familiar more broadly in systems with coupled heavy and light scalars, where violations of adiabaticity can lead to transient effects or invalidate naive low-energy descriptions \cite{Shiu:2011qw,Collins:2012nq}} The intermediate regime involving timescales comparable to $t_r$ is more complicated.

For the relaxation mechanism described above the characteristic scale is the Kaluza-Klein scale $m_\KK \sim 1/r$ and so $t_r \sim r$. Any process involving energies much larger than this should be regarded as fast for relaxation purposes while those involving energies much smaller than this are slow.  

To make these observations more explicit, we follow \cite{Burgess:2021obw,Burgess:2021juk} and add a cartoon of the Higgs potential into this framework through the choice
\be \label{WxHiggs}
  W_x \simeq m ( \phi - u) + \tfrac12 \lambda \Bigl( \cH^* \cH - v^2 \Bigr) \,,
\ee
since then the leading tension term in \pref{Expintau} becomes
\be \label{tausq2}
  V_2 = \frac{|W_x|^2}{\tau^2} = \frac{1}{\tau^2} \left[ m^2 |\phi - u|^2 + \tfrac14 \lambda^2 \Bigl(\cH^* \cH - v^2 \Bigr)^2 +  \lambda m\, \hbox{Re} ( \phi  - u)  \Bigl( \cH^* \cH - v^2 \Bigr) \right] \,.
\ee
This includes the mass term for $\phi$ and the standard Higgs potential plus a cross-term interaction that couples $\phi$ to $\cH$. 

The potential \pref{tausq2} has a trough-like shape whose minimum lies along the line 
\be
  m \phi = m u - \tfrac12 \lambda (\cH^* \cH - v^2) \qquad \hbox{(trough bottom)},
\ee
for all $\tau$. This guarantees that the matrix of second derivatives of the potential evaluated anywhere along this minimum (the `mass' matrix) has at least one zero eigenvector that is built from a combination of $\phi$ and $\cH$. For instance expanding $\cH = \ol{\cH} + \delta \cH$, $\phi = \ol{\phi} + \delta \phi$, $\tau = \tau_0 + \delta \tau$ and so on, where the background configuration satisfies $\ol{\cH}^* \ol{\cH} = v^2$ and $\ol{\phi} = u$ and $\tau_0$ arbitrary (but large) gives the following matrix of second derivatives
\be \label{massmatrix}
 \left( \frac{\partial^2 V_2}{\partial \psi^i \partial \psi^j} \right)_{\phi_0=u,\ol{\cH}=v,\tau_0} = \frac{1}{\tau_0^2} \left( \begin{matrix}   \lambda^2 v^2 &   0 &  \lambda m v &  0 & 0 \cr  0 &  0  &  0 &  0 & 0 \cr  \lambda m v &  0 & m^2 & 0 & 0 \cr  0 & 0 & 0 & m^2 & 0  \cr 0 & 0 & 0 & 0 & 0   \end{matrix} \right) 
 \ee
where $\psi^i$ collectively denotes the five real components $\{\cH_\ssR, \cH_\ssI, \phi_\ssR, \phi_\ssI , \tau\}$, where $\cH = \cH_\ssR + i \cH_\ssI$ and $\phi = \phi_\ssR + i \phi_\ssI$. 

This matrix has three trivial eigenvectors: $\delta \cH_\ssI$ and $\delta \tau$ (with eigenvalue zero) and $\delta \phi_\ssI$ (with eigenvalue $m^2$). The remaining two eigenstates are $\psi_+$ (with eigenvalue $m^2 + \lambda^2v^2$) and $\psi_0$ (with eigenvalue 0), and these are related to the original field fluctuations by 
\be \label{rotate}
  \left( \begin{matrix}  \psi_+ \cr  \psi_0    \end{matrix} \right) = \left( \begin{matrix}  \cos\vartheta &   \sin\vartheta  \cr - \sin \vartheta & \cos \vartheta  \end{matrix} \right)  \left( \begin{matrix}   \delta \cH_\ssR \cr \delta \phi_\ssR \end{matrix} \right) 
\ee
where 
\be \label{mixingangle}
  \tan \vartheta = \frac{m}{\lambda v} \,.
\ee
Of these $\delta \cH_\ssI$ is the usual Goldstone boson corresponding to phase rotations of $\cH$ that is typically gets eaten by the Higgs mechanism, while $\psi_0$ denotes the tangent to the bottom of the trough. 

\subsubsection*{Slow evolution}

Consider first the case where parameters are chosen so that $\lambda v^2 \ll m^2$ and where all fields are chosen to vary appreciably only over distance scales much larger than $m^{-1}$. In this case the mixing angle obtained from \pref{mixingangle} satisfies $\vartheta \simeq \frac{\pi}{2}$ and so $\psi_0$ is close to the $\delta \cH_\ssR$ direction while $\psi_+$ is dominantly $\delta \phi_\ssR$. 

In this case an approximate solution to the $\phi$ equation of motion is found by choosing $\phi(x) = \phi_m[\cH(x), \tau(x)]$, where $\phi_m$ is a value that solves $\partial V_2/\partial \phi = 0$ as a function of the other fields. This is an exact solution to the classical equations if $\cH$ and $\tau$ are themselves also chosen to be constants that lie on stationary points of the potential, but also remains a good approximate solution if $\cH$ and $\tau$ only vary slowly in space and time. In this case the derivative terms in the equation of motion for $\phi$ are negligible compared with the potential-energy penalty paid when deviating from the solution $\phi(x) = \phi_m[\cH(x), \tau(x)]$. 

This is the relaxation regime of interest to cosmology for which $\delta \phi_\ssR \simeq \psi_+$ is a heavy field that can be integrated out, leaving a residual potential for $\cH$ and $\tau$ that is much shallower (in this case precisely flat) than is the naive potential $\frac14\lambda^2 (\cH^* \cH - v^2)^2/\tau^2$ they would have had if $\phi$ had been held fixed at $\phi=u$. The corresponding masses are similarly suppressed (in this case they are zero), with the low-energy mass eigenstates given by the zero eigenvectors of \pref{massmatrix}. It is this regime whose cosmological implications are explored in detail below for the axion-dilaton system below.

\subsubsection*{Fast scattering}

Alternatively, next imagine choosing parameters where $\lambda v \gg m$, as would be appropriate (say) to Higgs physics at energies much higher than the relaxon mass. In this case the mixing angle obtained from \pref{mixingangle} satisfies $\vartheta \ll 1$ and so $\psi_0$ is close to the $\delta \phi_\ssR$ direction while $\psi_+$ is dominantly $\delta \cH_\ssR$. In this regime $\cH_\ssR$ is to good approximation a massive particle and $\phi_\ssR$ is dominantly very light, and so the field evolution is no longer well-approximated by the adiabatic choice $\phi(x) = \phi_m[\cH(x),\tau(x)]$. 

In this case particle scattering process are described by expanding $\cH = \ol{\cH} + \delta \cH$, $\phi = \phi_0 + \delta \phi$, $\tau = \tau_0 + \delta \tau$ as before, with the fluctuations describing wave packets 
\be
 \delta \cH = \int \frac{\exd^3k}{(2\pi)^3} \, f(\bfk) \, \exp[i \bfk \cdot \bfr - i \omega t] \,, 
\ee
with $\omega^2 = \bfk^2 + m^2_\ssH$, and similarly for $\delta \phi$ and $\delta \tau$. If these packets only interact over a short time in comparison to $t_r \sim m_\phi^{-1}$ and the other couplings (like $\lambda$) are weak then the future evolution can be computed perturbatively, leading to a standard scattering calculation involving Feynman diagrams computed (when $\tau_0$ is large) using the potential \pref{tausq2}. This potential is nontrivial because the fluctuating fields $\phi$, $\cH$ and $\tau$ do not minimize $V_2$ (although $\phi_0$ and $\ol{\cH}$ do). 

In this regime the mixing between $\cH_\ssR$ and $\phi_\ssR$ opens up an invisible width for the Higgs and allows processes where many light $\psi_0$ states are radiated by energetic $\cH$ fields, but without the Higgs field having its mass suppressed.  This is precisely what is required to capture the analogous processes in the UV description which involve a brane-localized Higgs radiating into the extra dimensions \cite{Burgess:2004yq, Atwood:2000au}. Notice that the presence of multiple fields in the bulk (as implied by supersymmetry) means these processes can be dominated by new types of portals \cite{Diener:2013xpa, Beauchemin:2004zi}, unlike the vanilla graviton emission usually studied in non-supersymmetric large-dimensional models \cite{Giudice:1998ck, Han:1998sg, Hewett:1998sn, Giudice:2003tu}. 

\section{Constraints from matter-dependent potentials}
\label{ssec:axionsinmasses}

The phenomenology of the models described in this paper depends in an important way on the axion's non-derivative couplings to matter, as we now argue. In some circumstances these can dominate derivative couplings like \pref{VanillaAxCoupling} discussed above, and this is true in particular when they induce a strong matter-dependent contribution to the axion scalar potential.  Most importantly, the coupling to gluons of a QCD axion is known to induce non-derivative axion-matter couplings within bulk matter in addition to the usual vacuum potential described earlier, and does so in a calculable way \cite{Cohen:1991nk}.

\subsection{Non-derivative axion-matter couplings}
\label{sssec:axionsinmasses}

For the purposes of relaxation the key assumption (as above) is that QCD is localized on a brane in the UV theory along with the rest of the Standard Model sector. As a result the matter couplings of the axion have the standard QCD form when expressed in terms of the $\tau$-independent Jordan frame metric $\tilde g_{\mu\nu}$ defined in \pref{JordanFrameDef}. This means they also inherit a $\tau$-dependence once one changes to the 4D Einstein-frame metric. Within the yoga framework there are two independent types of non-derivative axion-matter couplings that arise in this way, that we now summarize. 

\subsubsection{Universal bare mass shift}

The less important way that particles acquire non-derivative axion couplings is through an axion-dependent shift of the Higgs vev $v$. This shift can be seen by combining \pref{WxAxion} with \pref{WxHiggs} to get
\be \label{WxAxionHiggs}
  W_x \simeq m ( \phi - u) + \Lambda^2 \sin\left( \frac{\mfa}{2} \right) + \tfrac12 \lambda \Bigl( \cH^* \cH - v^2 \Bigr) \,,
\ee
since this generates Higgs-axion cross couplings once used in the leading potential $|W_x|^2/\tau^2$. When $\mfa$ is very slowly varying it can be treated as a background field when studying Higgs dynamics, and when this is true \pref{WxAxionHiggs} shows that the effective Higgs vev becomes 
\be
   v_\eff^2 = v^2 - \Lambda^2 \sin \left( \frac{\mfa}{2} \right) \,.
\ee

This axion-dependence is then inherited by all Standard Model bare masses since these are all proportional to $v_\eff$. This source of axion dependence is universal in the sense that it predicts all SM mass ratios are $\mfa$-independent. The relative size of this universal shift is order $\Lambda^2/v^2$, which for QCD implies an axion-dependent mass shift at the level $\delta m/m \sim 10^{-7}$. 

\subsubsection{QCD driven mass shifts}

For the QCD axion a larger contributor of axion-dependence to particle masses comes from QCD itself. This arises because the strong interactions contribute to the masses of any light composite particles (like protons or neutrons). The axion dependence of the mass acquired in this way can be calculated and takes the form
\be 
   - \frac{\cL_{\rm mass}}{\sqrt{-\tilde g}} = m_0 \Bigl[ 1 + U_{\rm mat}(\mfa) \Bigr] \ol \psi \, \psi \,,
\ee
where $\psi$ is a 4D spin-half field and the function $U_{\rm mat}$ is computed explicitly in \cite{Cohen:1991nk}. The result is particularly simple in the limit $m_u \simeq m_d$,
\be \label{QCDAxionMassContribution}
 U_{\rm mat}(\mfa)  \simeq\frac{ \sigma_\ssN}{m_0} \left[ \left| \cos \left(\frac{\mfa}{2} \right) \right| - 1 \right] 
 \quad \hbox{with} \quad \sigma_\ssN =  \sum_{q=u,d} m_q \, \frac{\partial m_\ssN}{\partial m_q}\,.
\ee
An additive constant has been added to ensure $U_{\rm mat}(\mfa = 0) = 0$ so that $m_0$ is the vacuum nucleon mass. Notice that $\mfa = 0$ is a local {\it maximum} of $U_{\rm mat}(\mfa)$ and so $U_{\rm mat}(\mfa)$ is not minimized by the same field configurations that minimize the vacuum potential \pref{QCDVacuumPotential}.\footnote{This fact is intriguing because it is a key ingredient of screening mechanisms like that discussed in \cite{Brax:2023qyp} that are designed to help hide the dangerous dilaton couplings from tests of GR.} 

Switching back to Einstein frame and keeping in mind the metric dependence of the $\psi$ kinetic term, this amounts to giving a $\tau$- and $\mfa$-dependent mass to composite baryons of size
\be
   m_\ssN(\mfa,\tau) = \frac{m_0}{\sqrt\tau} \Bigl[ 1 + U_{\rm mat}(\mfa) \Bigr] \,.
\ee
In Einstein frame the value of the nucleon mass implies $m_0/\sqrt\tau \simeq 940$ MeV and the value predicted for the axion-dependent coefficient is $\sigma_\ssN/\sqrt\tau \simeq 59$ MeV \cite{Alarcon:2011zs}, and so $\sigma_\ssN/m_0 \simeq 0.06$.  The $\tau$-dependence of this mass is not affected by relaxation ultimately for the reasons described in \S\ref{Fast vs Slow relaxation} above.

Within nonrelativistic bulk matter the presence of such a coupling turns the baryon energy density into a matter-dependent axion potential of the form
\be \label{VmatForm}
    V_{\rm mat}(\mfa,\tau) = \rho_\ssB(\mfa,\tau) = m_\ssN(\mfa,\tau) \, n_\ssB = \rho_{\ssB 0} (\tau)  \Bigl[ 1 + U_{\rm mat}(\mfa) \Bigr]
\ee
where $n_\ssB$ denotes the local baryon number density. This can lead to big effects when the baryon density is large enough to allow it to compete with the vacuum potential \pref{QCDVacuumPotential} \cite{Olive:2007aj,Banerjee:2025dlo}. In the usual telling of the QCD-axion tale the matter-dependent potential only competes with the vacuum potential 
for baryon densities $n_\ssB \gsim \Lambda_\QCD^4/m_\ssN$ and so only has interesting effects for nuclear densities (for an application of this observation see \cite{Hook:2017psm, LIGOScientific:2017ync}).

\begin{figure}[htbp]
    \centering
        \includegraphics[width=\linewidth]{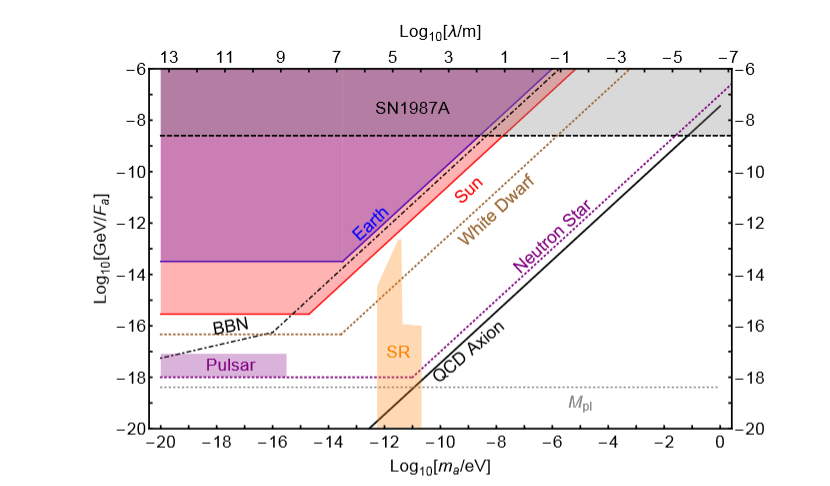}
 \hfill

    \caption{The region of the $1/F_a$-$m_a$ plane for which the QCD matter potential dominates the vacuum potential (within the Earth, Sun, white dwarves and neutron stars), for the vacuum-axion suppression model described in \cite{Hook:2017psm}.}
    \label{fig:MattVDominanceRegime}
\end{figure}

But this all changes once the vacuum potential is suppressed by a relaxation mechanism because then the relaxation of the vacuum potential allows the matter potential to dominate the vacuum potential for much smaller baryon densities. Quantitatively, the matter-dependent potential becomes competitive once $n_\ssB \gsim \Lambda_{\rm rel}^4/m_\ssN \sim \varepsilon^2/(m_\ssN r^4)$ rather than $\Lambda^4_{\rm unr}/m_\ssN \simeq \Lambda^4_\QCD/m_\ssN$. Using $1/r \sim 0.1$ eV and $\varepsilon = \Lambda_{\rm unr}/M_\EW \sim 10^{-5}$ (as appropriate for QCD) then shows that the matter potential dominates when $n_\ssB \gsim 10^{-23}$ eV${}^3 
\sim 10^{-3}$ per cubic metre. Because this is much smaller than the typical matter densities of everyday experience (or of late-time cosmology) whenever \pref{VmatForm} is a good approximation we should expect it to dominate the QCD vacuum potential.

\subsubsection{Mass-dependence constraints}

Very strong constraints can arise if the QCD matter-dependent potential \pref{VmatForm} should dominate the vacuum potential \pref{QCDVacuumPotential} within ordinary matter \cite{Hook:2017psm}, some of which are shown in Fig.~\ref{fig:MattVDominanceRegime}. A common feature of these -- visible in the figure -- is that they only apply to the left of the naive QCD-axion line. It is tempting to think that this means they do not constrain models (like the bulk axion model described here) lying to the right of the QCD line, but this can be deceptive because the constraints really depend on the value of $\Lambda_{\rm rel}$ and on how $F_a$ is related to the axion kinetic coefficient $f$, which we've seen can change in relaxation models.

In particular, the bounds of \cite{Hook:2017psm} are derived within the `sudden' approximation, which assumes the transition from matter-potential dominance to vacuum-potential dominance occurs over length scales $\ell \sim n/(\exd n/\exd r)$ that are much smaller than the axion's Compton wavelength $m_{\rm in}^{-1}$ inside matter. Under this assumption the axion sits at the minimum $\mfa \neq 0$ of the matter potential throughout an object's interior, matching on to a Yukawa-type falloff in the vacuum-dominated exterior region. The strong bounds then emerge from the absence of evidence for changes to nuclear physics (such as the pion mass, the proton/neutron mass difference or the emergence of neutron electric dipole moments) that occur within the object's interior and near its surface when the background axion is not at $\mfa = 0$ \cite{Hook:2017psm, Ubaldi:2008nf, Lattimer:2012nd, Mukai:2017qww, Audi:2003zz}. Many (but not all) of these bounds also apply in the opposite `adiabatic' approximation where $\ell \gg m_{\rm in}^{-1}$ for which the interior axion field simply tracks the minimum of the summed vacuum and matter potentials and so emerges at the surface with $\mfa = 0$ already achieved (see \cite{Brax:2023qyp} for a discussion of this adiabatic regime). 

This leaves the question of when the matter potential \pref{VmatForm} is a good approximation to use when describing axion profiles within matter. And when does this matter potential dominate over the vacuum potential? The latter question is the easier to answer: the matter potential dominates the vacuum potential when $\rho_{\rm mat} \simeq m_\ssN n_\ssB > \Lambda_{\rm rel}^4 = m_a^2 f^2$, showing that the matter potential dominates when $1/f \gsim m_a/\sqrt{\rho_{\rm mat}}$. This reproduces the condition $\rho_{\rm mat} > m_a^2 F_a^2$ used in \cite{Hook:2017psm} (though with $f$ appearing rather than $F_a$) to obtain the diagonal boundary to the lower-right of the constrained regions shown in Fig.~\ref{fig:MattVDominanceRegime}.

The horizontal lower boundary of the excluded region in this figure is found using the observation that large density $n > \Lambda_{\rm rel}^4/m_\ssN$ is not in itself sufficient to displace the axion from its vacuum minimum; the size of the high-density region also matters \cite{Khoury:2003aq,Hook:2017psm,Burrage:2021nys,Balkin:2022qer}. An estimate of the minimum size $L$ required to allow the interior axion to differ appreciably from its vacuum value can be obtained by asking the potential energy gain to be much larger than the gradient energy required to return to the vacuum value outside. Within the sudden approximation this requires $(m_\ssN n_\ssB  - \Lambda_{\rm rel}^4 )L^3 \simeq m_\ssN n_\ssB L^3 \sim \rho_{\rm mat} L^3$ to be much larger than the gradient energy $(f^2/L^2) L^3 \sim f^2 L$ cost, showing that there is a minimum system size 
\be \label{f vs rho L}
   f \lsim \sqrt{\rho_{\rm mat}} \; L \,,
\ee
below which the axion is not well-approximated by the constant field that minimizes the matter or matter plus vacuum potential. 

For brane-localized axions  we have $F_a \simeq f$ and so \pref{f vs rho L} can be read as a lower bound on the coupling $1/F_a$ once $\rho_{\rm mat}$ and $L$ are specified for an object, leading to the same constrained regions as are given in \cite{Hook:2017psm} and shown in Fig.~\ref{fig:MattVDominanceRegime}. As we've seen though, relaxation moves the brane-QCD line to the left of the naive line (see Fig.~\ref{fig:fa vs mass w yoga}) and so the relaxed QCD (brane) axion is pushed into a highly constrained region: both non-derivative and derivative couplings to matter rule out axions in this regime.

The situation is different for the bulk axion, for which \pref{f vs rho L} becomes
\be
 f \sim \frac{f_\ssB^2}{\tau F_a} \sim  \frac{M_p}{\tau} \sim \frac{1}{r}  \lsim \sqrt{\rho_{\rm mat}} \; L 
 \qquad \hbox{(bulk).} 
\ee
Unlike for either the brane-relaxed or naive QCD axion, this inequality is already satisfied within ordinary atomic nuclei, for which $\rho_{\rm mat} \sim \Lambda_\QCD^4$ and $L \sim \Lambda_\QCD^{-1} A^{1/3}$ (where $A$ is the nuclear atomic number) together imply $\sqrt{\rho_{\rm mat}}\, L \sim \Lambda_\QCD A^{1/3} \sim 10^8 A^{1/3}$ eV, which is much larger than $1/r \sim 0.1$ eV. Having $\mfa \neq 0$ inside a nucleus is likely also to be phenomenological poison, since it re-opens the possibility of large neutron electric dipole moments and other unacceptable changes to nuclear physics (such as to the neutron-proton mass difference). 

The detailed phenomenological implications of this regime likely require more care because the nuclear size is much smaller than the extra-dimensional size and so the axion response is better computed using the full extra-dimensional field equations, along the lines done in these models for the dilaton and metric in \cite{Burgess:2014qha}.
In what follows we leave open whether the late-time couplings to matter of the axiodilaton can be different from those relevant to cosmology and so set aside a more detailed discussion of the present-day phenomenology of the bulk axion. We instead next focus on its cosmological properties.

\subsection{Constraints from cosmology}
\label{sec:LEDynamics}

Cosmology is the perfect gibbet on which to hang modification to the standard story of QCD axion physics. The gravitational implications of light scalars are inescapable across these length scales, and the dynamics are forced to play out in a setting where the relevant sources are both homogeneous and time--dependent (and so particularly simple)  to a good approximation. In compact astrophysical systems one can often hide behind environmental complexity of many body systems and uncertain profiles interior to the systems \cite{Jain:2012tn,Chan:2021bja,Tamosiunas:2021kth}; in cosmology the mean densities and the expansion history are fixed functions of time, so any matter--dependent contribution to the axion sector is driven by a known and steadily changing background. 

This makes the usual coarse--graining unusually reliable (at least up until the relatively recent past when baryons start to clump significantly), thanks to the enormous separation between microphysical and Hubble scales, tying background evolution to standard early--Universe anchors in a way that leaves little room for loopholes. We therefore here ask whether relaxed QCD axions can provide a viable description of the physics we know and love on large scales \emph{anywhere} in the Universe, independent of systematics endemic to compact astrophysical environments.

The logic of relaxation provides an additional motivation for focusing on cosmology since these mechanisms preferentially affect slow processes while leaving faster laboratory and particle--physics processes comparatively untouched. Cosmology, being the slowest evolution one can possibly probe, is therefore the regime in which any distinctive behaviour should be most apparent. 

Along the bottom of the trough the dilaton $\tau$ evolves on Hubble timescales, while the axion
dynamics are much faster. This separation of scales allows us to treat $\tau$ as an
external, slowly varying parameter when analysing the axion potential. As discussed around
\pref{axio-dil pot0}, once the relaxon has been minimized the important point is not the
detailed trigonometric form of the residual axion dependence, but rather that the vacuum
potential along the trough remains periodic in $\mfa$ and is suppressed by the relaxed scale
$\varepsilon^2 V_0/\tau^4$. Motivated by this, and by the QCD discussion around
\pref{QCDVacuumPotential} and \pref{QCDAxionMassContribution}, we write the vacuum and
matter contributions in the generic form
\be
  V_{\rm vac}(\mfa,\tau)
  \;=\;
  -\,\varepsilon^2\,\frac{V_0}{\tau^4}\,u(\mfa)\,,
  \label{periodic_cosmological_pot}
\ee
and
\be
  V_{\rm mat}(\mfa,\tau)
  \;=\;
  \rho_b\,\frac{\sigma_{\ssN}}{m_0}\,u(\mfa)\,,
  \label{eq:Vmat_generic}
\ee
where $u(\mfa)$ is a dimensionless periodic order-unity function. The QCD expressions above
motivate taking the matter-induced contribution to have the same shape as the vacuum potential
but with the opposite sign, up to an irrelevant additive constant that has been absorbed into the
background energy density. The two contributions therefore have extrema at the same field values, but
their roles are reversed: a minimum of the vacuum potential coincides with a
maximum of the matter potential, and vice versa. We denote by $\mfa_+$ a
CP-conserving vacuum minimum and by $\mfa_-$ the neighbouring matter-selected
minimum. For the QCD-motivated periodicity discussed around
\pref{QCDVacuumPotential} and \pref{QCDAxionMassContribution}, these occur at
$\mfa_+=4n\pi$ and $\mfa_-=(2n+1)2\pi$.

The axion effective potential in a homogeneous baryon background is then
\be
  V_{\rm eff}(\mfa,\tau)
  \;=\;
  V_{\rm vac}(\mfa,\tau)+V_{\rm mat}(\mfa,\tau)
  \;=\;
  \left[
    \rho_b\,\frac{\sigma_{\ssN}}{m_0}
    -
    \varepsilon^2\,\frac{V_0}{\tau^4}
  \right] u(\mfa)\,.
  \label{eq:Veff_simple}
\ee
The stationary points are therefore determined simply by $u'(\mfa)=0$, and which of these
stationary points is a minimum depends only on the sign of the overall coefficient. The minimum transitions from $\mfa_-$ to $\mfa_+$ when the density is near a threshold density defined by the ratio of strengths of matter and relaxed vacuum potentials:
\be  \label{eq:rho_crit_def}
  \rho_\ssB^{\rm th} =  \frac{m_0 V_0 \varepsilon^2}{\sigma_\ssN\tau^4} \simeq 17 \, \rho_{\DE} \, \varepsilon^2 \simeq 240 \, \rho_{\ssB 0} \, \varepsilon^2 \,.
\ee
Here the numerical estimate uses $\sigma_\ssN/m_0 \simeq 0.06$ as above and the fact that $V_0/\tau^4$ is by construction of order the present-day Dark Energy density $\rho_\DE \sim  10^{-30} \,{\rm g\,cm^{-3}} \simeq 10^{-10}$ eV${}^4$. The final equality uses $\rho_\DE = \rho_{\ssB 0}(\Omega_\DE/\Omega_{\ssB 0})$ where today's cosmic mean baryon density is $\rho_{\ssB 0}\sim 10^{-31}\,{\rm g\,cm^{-3}} \sim 10^{-11}$ eV${}^4$ while $\Omega_{\ssB 0} \simeq 0.05$ and $\Omega_\DE \simeq 0.7$ are the current baryon and Dark Energy mass fractions. Notice that this critical density is similar for both the brane and bulk axion scenarios because it does not depend on the axion kinetic normalisation.

When interpreting $\rho_\ssB^{\rm th}$ it is useful to keep in mind the typical size of the mean density of baryons at different epochs. As mentioned above, the cosmic mean baryon density today is $\rho_{\ssB 0}\sim 10^{-31}\,{\rm g\,cm^{-3}} \sim 10^{-11}$ eV${}^4$, rising to $\rho_\ssB(z_{\rm eq})\sim 10^{-20}\,{\rm g\,cm^{-3}} \simeq 1$ eV${}^4$ at matter--radiation equality, whereas typical terrestrial and stellar interior densities are much larger: $\rho_\ssB\sim 1$--$10^2\,{\rm g\,cm^{-3}} \sim 10^{20}$--$10^{22}$ eV${}^4$ (and nuclear densities are larger again).

The predictions of cosmology for the potential are then both swift and sure because $\rho_\ssB(z)$ is relatively well understood, evolving with redshift as $\rho_\ssB(z)\approx\rho_{\ssB 0}(1+z)^3$, where $\rho_{\ssB 0}$. It in particular equals the critical density at a redshift  
\be   \label{eq:zeq_eps}
  z_{\rm th} =
  \left(\frac{\rho_\ssB^{\rm th}}{\rho_{\ssB 0}}\right)^{\!1/3} - 1 \simeq \cO(10^2)\,\varepsilon^{2/3}-1.
\ee
This is order unity when $\varepsilon \sim 10^{-3}$ and so is negative for the QCD--motivated choice $\varepsilon\sim10^{-5}$, implying $\rho_\ssB(z)>\rho_\ssB^{\rm th}$ throughout the entire past history of the Universe. Once the vacuum term is relaxed into the $\tau^{-4}$
trough, even the extremely dilute cosmological background remains above the threshold (for QCD-like parameters).

When the evolution of the potential due to cosmological expansion is very slow relative to the axion's oscillation frequency, the axion evolution is adiabatic, with the homogeneous field well-approximated by the value of the minimum of \pref{eq:Veff_simple} transitioning between $\mfa_+$ and $\mfa_-$. To determine whether the evolution is adiabatic requires knowing the axion mass, and this is where brane and bulk axions could in principle differ because their kinetic prefactors differ in their $\tau$--dependence, with
\be   \label{eq:kin_scaling_n}
  -\frac{\cL_{\rm kin}}{\sqrt{-g}}
  \;=\;
  \frac{M_p^2}{2}\,\tau^{-n}\,\partial_\mu\mfa\,\partial^\mu\mfa,
  \qquad
  n=
  \begin{cases}
    1 & \text{brane axion},\\
    2 & \text{bulk axion},
  \end{cases}
\ee
so that the dynamical oscillation frequency about the minimum is set not by
$V_{{\rm eff},\mfa\mfa}$ alone, but by the kinetic--weighted curvature. For the
QCD--motivated periodic dependence discussed earlier the axion appears inside
trigonometric functions whose arguments
are $\cO(1)$. Their derivatives therefore introduce
only order--unity numerical factors, so that the periodic function $u(\mfa)$ and its
second derivative both remain $\cO(1)$. The effective curvature is therefore
\be
  m_{\mfa,{\rm eff}}^2
  \;:=\;
  \frac{\tau^{n}}{M_p^2}\,\pdv[2]{V_{\rm eff}}{\mfa}
  \;\sim\;
  \frac{\tau^{n}}{M_p^2}\!
  \left|
  \rho_b\,\frac{\sigma_{\ssN}}{m_0}
  -
  \varepsilon^2\,\frac{V_0}{\tau^4}
  \right|,
  \label{eq:meff_total}
\ee
up to order--unity factors set by the precise periodic shape.

The adiabatic approximation requires $m_{\mfa,{\rm eff}}\gg H$ so that the minimum drifts slowly compared to $m_{\mfa,{\rm eff}}^{-1}$. In the trough regime both conditions are satisfied in the late Universe for the benchmark masses derived earlier: for $\tau \sim 10^{28}$ the relaxed axion mass is $m_\mfa\sim 10^{-20}\,{\rm eV}$ for brane axions and $m_\mfa\sim 10^{-5}\,{\rm eV}$ for bulk axions (see the mass estimates around \pref{braneFa} and \pref{fmsizebulk}). By comparison, $H_0\sim 10^{-33}\,{\rm eV}$ today and $H(z_{\rm eq})\sim 10^{-28}\,{\rm eV}$ at matter--radiation equality, so both types of axions are safely adiabatic with average values that track the minimum of \pref{eq:Veff_simple}. 

It follows that the mean axion field is driven to $\mfa_-$ for much of the universe's history, so long as $\rho_\ssB>\rho_\ssB^{\rm th}$. If the axion also couples to the dominant cold dark matter density (rather than baryons only), the replacement $\rho_\ssB\to\rho_m$ strengthens the conclusion by an additional factor
$\rho_m/\rho_\ssB\simeq\Omega_m/\Omega_b$.

For the very recent universe the clumpiness of the baryon distribution makes suspect the mean-field approximation wherein the baryon density is replaced by its average, and a more careful inhomogeneous treatment of axion evolution is required. That is all the more true when the kinetic coefficient is small since then the penalty for spatial gradients is less expensive to pay. 

But these complications are unlikely to matter for redshifts larger than about 10, at which times the axion can be expected to be nearer the matter--selected extremum $\mfa\simeq\mfa_-$ than the CP-conserving vacuum minimum $\mfa\simeq\mfa_+$. This is a serious problem for phenomenology for several reasons. First, if the average axion configuration is not near its CP-conserving minimum even now then analyses that track how the axion evolves with position near macroscopic bodies must rethink their cosmological boundary conditions. 

One might entertain taking the vacuum potential to be much larger, such as if $\varepsilon\gtrsim 10^{-3}$, so that the present-day mean axion can be acceptable (though in this case the model must then part company from the QCD scale by many orders of magnitude). But even then phenomenology remains a problem because in general nuclear properties at BBN and recombination would be unacceptably different than they are today \cite{Dent:2007et,Lee:2020tmi}. 

We can therefore safely conclude that once the axion vacuum potential is relaxed into the $\tau^{-4}$ trough, the cosmological mean baryon density very likely lies above the critical density \pref{eq:rho_crit_def} if its unsuppressed size is set by the QCD scale. In such scenarios the matter potential dominates axion evolution for most of the universe's history and its mass is large enough that its mean value generically adiabatically tracks the minimum $\mfa_-$ of the matter potential rather than the CP--conserving vacuum minimum $\mfa_+ = 0$. Because the axion is not Hubble--frozen after matter--radiation equality (for either the brane or bulk kinetic scalings), it has ample time to track this minimum throughout the late Universe.

The same would be expected to be true in spades near compact objects, like galaxies, stars and planets, whose densities sit many orders of magnitude deeper in the $\rho_\ssB\gg\rho_\ssB^{\rm th}$ regime. It might be possible for the vacuum minimum to dominate the matter-dependent one (such as the most underdense cosmic voids), but these occupy such a small fraction of the mass budget that they do not provide a generic cosmological attractor for the axion field. 

We conclude that a yoga-style relaxation mechanism for suppressing the Dark Energy (and very likely other mechanisms as well) does not accommodate an axion solution to the strong--CP problem.

\section{Conclusions}
\label{sec:Conclusions}

This paper uses natural relaxation as a concrete framework for illustrating a broader point.
Any mechanism that dynamically suppresses the vacuum energy is generically not confined to the
vacuum term alone. If the suppression is achieved by adiabatic tracking of light fields, then
other scalar potentials that are slowly varying on cosmological timescales are also reshaped.
The QCD axion provides a particularly sharp test case because its standard phenomenology
relies on a specific relation between its potential, its mass, and its couplings and because QCD
also fixes a calculable dependence of hadronic masses on the axion background.

\subsection{Lessons learned}

Embedding a Peccei--Quinn axion into the yoga framework makes the axion vacuum potential inherit
the same suppression that is required for Dark Energy. The resulting axion mass--coupling
relation is therefore displaced from the usual QCD band. In the simplest brane realisations this
displacement pushes the axion into regions of the coupling--mass plane that are already strongly
constrained by existing searches, and it simultaneously undermines the standard misalignment
mechanism for producing axion dark matter. Dual or bulk realisations can evade some of these
direct constraints because the relation between the axion kinetic normalisation and the matter
coupling is altered by duality, but this does not make the axion safe.

The more robust obstruction arises from the interplay between vacuum and matter potentials once
the vacuum term is suppressed. QCD implies a non-derivative axion dependence of baryon masses,
and in bulk matter this becomes a matter-induced contribution to the axion effective potential.
In conventional axion physics this contribution is only competitive at very high densities.
In relaxation frameworks the vacuum potential is parametrically smaller, so the matter-induced
term typically dominates already at ordinary densities and, crucially, at the cosmological mean
baryon density. The result is that, for QCD-motivated parameter choices, the axion is
forced toward the matter-selected extremum rather than the CP-conserving vacuum minimum
throughout the post--BBN history and even into the far future. Avoiding this requires pushing the
parameters into a regime that no longer resembles a Peccei--Quinn QCD axion.

A relaxation mechanism that succeeds at suppressing the
vacuum energy generically spoils the standard QCD axion in its usual form, either by moving it
into already excluded regions of parameter space or by making its cosmological evolution select
the wrong minimum once matter effects are included. At the same time, the framework also clarifies
why this conclusion does not automatically extend to all of particle physics. Relaxation is a
dynamical response that is efficient only for slow evolution. Fast, high-energy processes such as
those probed in collider Higgs physics occur on timescales too short for the relaxation fields to
track, so they remain well described by the usual unsuppressed potentials. The tension is instead
sharpest for light scalars whose dynamics unfold on Hubble timescales, precisely the regime in
which cosmology provides the cleanest laboratory.

\subsection{Future work}

On the model-building side the most interesting open direction is to understand what ingredients,
if any, can preserve a viable axion while retaining the benefits of vacuum-energy relaxation.
This could involve embeddings in which the axion potential is generated but with a minimum that is shared by both matter and vacuum potentials. Or the axion could be in a sector that couples to matter only through field derivatives, since in this case no matter potential need be generated at all. Alternatively, perhaps mechanisms can be found that suppress or screen the
relevant non-derivative axion couplings inside matter without simultaneously undoing the Peccei--Quinn
solution to strong CP.

Closely related to this is the unresolved issue of screening the matter-dilaton interactions (though hopefully doing so without ruining the alleviation of the Hubble tension to which sizable dilaton-matter couplings can lead \cite{Smith:2025uaq}). The yoga framework
predicts a light scalar with gravitational-strength couplings to matter, and phenomenological viability
likely requires an additional mechanism that reduces these couplings in the late Universe \cite{Brax:2023qyp,Brax:2026ttj}. A complete
picture must track how any such mechanism feeds into axion couplings and into the matter-induced
potential that drives the cosmological no-go identified here.

More broadly, we have used the Peccei--Quinn axion as one representative example of a larger
class of light scalars that are often invoked in the Universe to solve long-standing problems, like dark matter, dark energy, or other
components of the dark sector \cite{Arvanitaki:2009fg,vandeBruck:2022xbk,Ferreira:2020fam,Poulin:2023lkg}. The analysis here suggests that confronting the cosmological
constant problem can place powerful and sometimes unexpected constraints on such degrees of
freedom. It also highlights a potential tension for scalar--field dark matter within relaxation
frameworks. Realising dark matter typically requires an oscillating field with a mass scale much
larger than the QCD scale relevant for the axion sector studied here \cite{Turner:1983he,Eberhardt:2025caq,Wardana:2026qag}. In a theory where the
relaxation mechanism relies on a shallow dilaton direction to maintain a small vacuum energy,
a heavier and more energetic oscillating scalar can backreact on the slowly varying background (as has been observed more broadly in e.g. \cite{Marsh:2011gr,Marsh:2012nm,Alam:2022rtt}).
When the field oscillates with order--one excursions from its minimum, its time--dependent
energy density and induced couplings can periodically shift the location and even the existence
of the effective dilaton minimum, potentially destabilising the late--time trough solution.
Determining when this occurs requires a dedicated analysis that keeps track of the oscillation
amplitude, the fraction of the cosmic energy density stored in the scalar, and the induced
modulation of the effective potential.

\section*{Acknowledgments}
\noindent We thank Marco Costa, Michael Fedderke, Junwu Huang and Zach Weiner for helpful discussions. CvdB is supported by the Lancaster–Sheffield Consortium for Fundamental Physics under STFC grant: ST/X000621/1. CB's research was partially supported by funds from the Natural Sciences and Engineering Research Council (NSERC) of Canada. AS is supported by the W.D. Collins Scholarship. AS is also grateful for the hospitality of Perimeter Institute and Kavli IPMU where part of this work was carried out. Research at the Perimeter Institute is supported in part by the Government of Canada through NSERC and by the Province of Ontario through MRI.

\appendix

\bibliographystyle{JHEP}
\bibliography{bibliography}

\end{document}